\documentclass[%
 preprint, linenumbers,
 amsmath,amssymb,
 aps, physrev,
]{revtex4-2}

\usepackage{graphicx}% Include figure files
\usepackage{dcolumn}% Align table columns on decimal point
\usepackage{bm}% bold math
\usepackage{color}
\usepackage[normalem]{ulem}
\usepackage{lineno}
\begin{document}
\nolinenumbers

\preprint{APS/123-QED}

\title{\textbf{Dynamics and universal scaling of Worthington jets in the cavity-free regime} 
}% 

\author{Xingsheng Li}
\author{Jing Li}%
 \email{Contact author: lijing\_@sjtu.edu.cn}
\affiliation{%
Marine Numerical Experimental Center, State Key Laboratory of Ocean Engineering, School of Ocean and Civil Engineering, Shanghai Jiao Tong University, Shanghai, 200240, PR China
}%


\date{\today}% It is always \today, today,

\begin{abstract}

Worthington jets ejected after the impact of a solid or liquid object on a liquid surface have extensive applications in natural, industrial, and scientific contexts.
Here, we present a combined experimental and theoretical investigation of the jet generated by sphere impact with no cavity formed.
Experiments identify three distinct pinch-off modes, whose regime boundaries are independent of sphere wettability and density, and are theoretically determined by the Rayleigh--Plateau instability.
From momentum and energy conservation, a new scaling law is derived for the dimensionless maximum jet height and agrees remarkably well with experiments across various impact conditions, thus validating its universal character and clarifying its dependence on the Froude, Weber, and Reynolds numbers as well as the density ratio.
Coupling self-similar solutions with a kinematic condition at the jet tip yields good predictions for the evolution of jet height and shape, revealing gravity-dominated jet dynamics, with a modification from surface tension that is most pronounced without pinch-off.
These findings demonstrate that the upward jet is sustained by the collision of converging flow behind the sphere, a generation mechanism fundamentally distinct from the cavity collapsing forced case.

\begin{description}

\item[Usage]
Secondary publications and information retrieval purposes.

\end{description}

\end{abstract}

\maketitle

%\tableofcontents

\section{\label{sec:Introduction}Introduction}

Worthington jets generated during the free surface impact of solid or liquid objects have diverse beneficial applications, ranging from human Olympic diving \citep{rubin1999basics} and droplet manipulation \citep{PhysRevApplied.3.044018} to spray cooling and painting \citep{AZIZ20002841}.
At the same time, their uncontrolled occurrence can result in serious pesticide wasting \citep{Nuruzzaman}, air pollution \citep{landrigan2018lancet}, and even the spread of pollutants and pathogens \citep{joung2015aerosol}.
Given these dual influences, this impact problem has attracted sustained interest since Worthington's pioneering work more than a century ago \citep{worthington1897v}.
Extensive research has examined jets produced by various impact objects, including liquid droplets \citep{marcotte2019ejecta,kim2021impact,van2021self,dong2023pinching}, solid spheres \citep{grumstrup2007cavity,duclaux2007dynamics,aristoff2009water,truscott2012unsteady,speirs2019water}, and flat disks \citep{bergmann2006giant,gekle2010supersonic,gekle2010generation}.
Similar phenomena have also been well investigated in the context of bubble bursting near a liquid surface \citep{lai2018bubble,kang2019gravity,dubitsky2023enrichment}.

Upon the impact of a solid body on a liquid surface, extensive research has focused on the air entrainment circumstance.
In addition to the classical axisymmetric cavity, various complex phenomena have been reported, including the appearance of asymmetric cavities for spheres with heterogeneous wetting properties \cite{watson2021water}, the formation of subsurface cavities and small droplets together with bubbles on the sphere surface due to the presence of a bubble layer resting on the surface of a water--surfactant mixture \cite{speirs2018entry}, and the development of a three-dimensional crumpled morphology along the cavity wall resulting from the existence of an oil lens floating on the water \cite{smolka2019sphere}.
As the axisymmetric air cavity collapses due to hydrostatic pressure with a local scaling exponent $\gamma = d \ln{h_0/d} \ln{t'}$ of the minimum radius $h_0$ of the cavity \citep{eggers2007theory}, the pinch-off eventually occurs at a single point, from which two jets are ejected upward and downward, respectively \citep{gekle2010generation}.
The upward Worthington jet has been verified to be driven by local flow around the base of the jet, which is forced by the colliding cavity wall, as explained by the theory of a collapsing void \citep{gekle2009high}.
The Rayleigh–Plesset equation has been widely and effectively applied to model such cavity collapse dynamics \citep{duclaux2007dynamics,aristoff2009water,plesset1977bubble}.
In particular, \citet{aristoff2009water} developed a theoretical model to describe the evolution of cavity shape and obtained an analytical solution for pinch-off events, which shows excellent agreement with the experiments.
They also reported that the cavity exhibits different sealing behaviors depending on the impact parameters, an observation that was later extended by \citet{speirs2019water} to incorporate the effect of sphere wettability.
Although this theory accurately captures cavity dynamics, it provides limited direct insight into the jet.
Alternatively, boundary integral simulations provide another powerful approach to the cavity collapse problem and have advanced the physical understanding of associated jet formation \citep{bergmann2009controlled,gekle2010generation}.

Above the free surface, considerable attention has been paid to the splash crown, a visually striking and dynamically rich phenomenon that continues to attract scientific interest.
\citet{marston2016crown} observed a buckling-type azimuthal instability of the crown, manifested as vertical striations.
\citet{eshraghi2020seal} later developed a theoretical model to describe the trajectory and dynamics of the crown, revealing its dependence on cavity pressure difference, gravity, and surface tension.
In the study of the effect of ambient gas on cavity formation, \citet{williams2022effect} found that the presence of gas delays crown sealing, thus affecting the condition for cavity formation.

In contrast to the extensively studied cavity collapse and its related phenomena, analysis of the sphere impact problem with no cavity formed remains comparatively limited.
A key theoretical contribution was made by \citet{duez2007making}, who derived the critical condition for cavity formation.
When no cavity is formed during the impact process, the jet generation mechanism is fundamentally different.
As illustrated in Figs.~\ref{Fig:snapshots}(a--c), the kinetic energy of the sphere is transferred directly to the surrounding liquid upon impact, providing the primary momentum to support the generation of the jet.
In this regime, traditional models based on the Rayleigh--Besant problem are no longer applicable, and attention to this kind of jet is relatively sparse.
Existing research has established only a basic relationship from an energy perspective, reporting that the maximum jet height is proportional to the release height \citep{Watson2018}.
Nevertheless, the detailed behavior of the jet, the parameters that govern the jet dynamics, and the underlying physical mechanisms have not yet been adequately clarified or explained.
Despite limited prior studies, insights can be drawn from droplet impact problem \citep{michon2017jet,C9SM00318E,PhysRevE.82.036319,PhysRevE.92.053022,10.1063/5.0084456}, where the resulting jet shape closely resembles the present observations (Figs.~\ref{Fig:snapshots}(a--c)).
Nevertheless, the underlying generation mechanisms differ fundamentally.
In droplet impact, although the formed crater does not undergo the same pinch-off collapse, as does the cavity caused by traditional sphere impact, the resulting upward jet is still driven by crater retraction, a process clearly shown in Ref.~\citep{PhysRevE.92.053022}.
As demonstrated in Refs.~\citep{michon2017jet,PhysRevE.82.036319,C9SM00318E}, energy is accumulated and transferred to the jet via capillary waves that propagate and focus at the bottom of the receding crater, ultimately giving rise to a vertical jet shooting out.

In the present work, we aim to systematically investigate Worthington jets generated by sphere impact on a liquid surface with no cavity formed, hereafter referred to as cavity-free jets.
In Sec.~\ref{sec:Experimental procedure}, the experimental procedure and the tested impact conditions are introduced.
Section~\ref{sec:Qualitative analysis} presents a qualitative analysis of the observed jet behavior, reports a theoretical explanation for the identified pinch-off regimes, and a brief discussion of the critical condition for cavity formation.
Section~\ref{sec:Jet dynamics} provides a quantitative analysis of jet dynamics, including the derivation of a scaling law for the maximum jet height and the application of a theoretical model to predict the evolution of jet height and shape.
Comparison with experimental results validates these frameworks.
Finally, conclusions are summarized in Sec.~\ref{sec:Conclusions}.

\section{\label{sec:Experimental procedure}Experimental procedure}

\begin{figure}
\centering
\includegraphics[width=460pt]{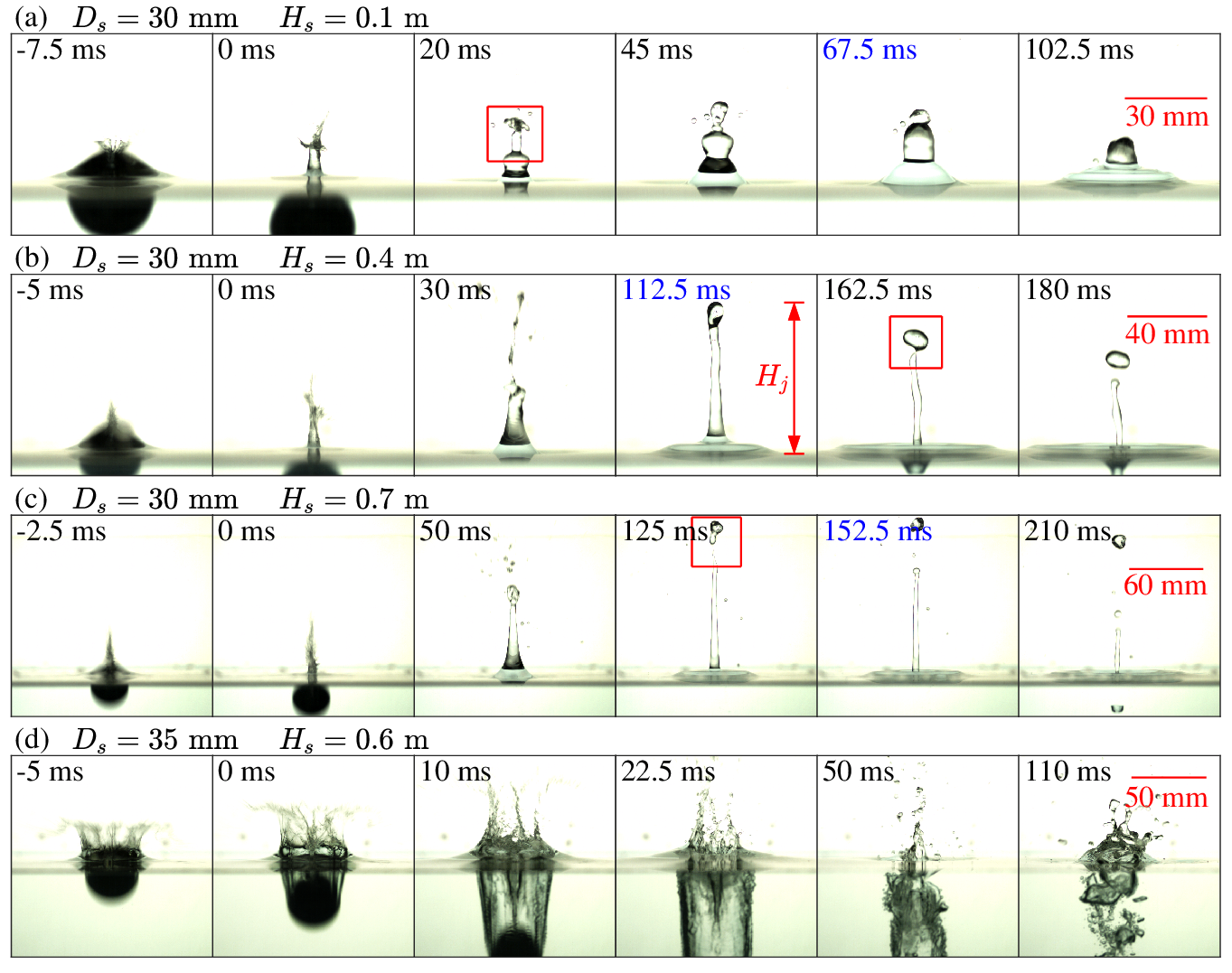}
\caption{\label{Fig:snapshots}Temporal evolution of Worthington jets or cavity dynamics following the impact of steel spheres on a water surface.
Rows (a--c) show the jet evolution for spheres of diameter $D_s=30\ \mathrm{mm}$ at release heights $H_s=0.1\ \mathrm{m},\ 0.4\ \mathrm{m},\ 0.7\ \mathrm{m}$, respectively.
Row (d) shows cavity dynamics for a hydrophobic-coated sphere of $D_s=35\ \mathrm{mm}$ at $H_s=0.6\ \mathrm{m}$.
Snapshots with blue timestamps indicate the instant when the jet reaches its maximum height.
Red boxes highlight key features: the formation of a fine splash at the initial stage of the jet in (a), and the pinch-off of the liquid column in (b) and (c).
The experimental GIF animations corresponding to cases (a)--(c) are provided in the Supplemental Material~\citep{supplemental2025}.}
\end{figure}

In our experiments, a solid sphere is initially held by a pneumatic finger at a specified height $H_s$ above the liquid free surface.
Each test is performed by releasing the sphere into a glass tank (0.5 m $\times$ 0.5 m $\times$ 0.5 m) filled with the working liquid.
The impact velocity is estimated as $U_s\approx \sqrt{2gH_s}$ under free-fall motion, where $g$ is the gravitational acceleration.
The subsequent evolution of the free surface is recorded by a high-speed camera at 400 frames per second, a frame rate sufficient to resolve the jet dynamics.
With a spatial resolution of 1696 $\times$ 1710 pixels ($\sim$ 0.126 mm/pixel), the camera accurately captures both the jet height and its shape.
A light source combined with a diffuser screen is utilized to ensure uniform illumination intensity and high image quality.
To guarantee data reliability and minimize experimental uncertainty, each parameter configuration is tested with at least 3 repeated tests, and the results are averaged.
The time origin $t=0$ is defined as the instant when the sphere becomes completely submerged.
Figures~\ref{Fig:snapshots}(a--c) illustrate the evolution of the jet after the impact of a sphere on the water surface.
Shortly after impact, a fine splash first forms at the jet tip.
After tens of milliseconds, this splash detaches from the main liquid column and breaks up into tiny droplets.
Because these splashes and droplets in the early stage of the jet are sufficiently tiny, they are not considered in the analysis of jet shape and height.

To systematically vary the experimental conditions, four types of spheres made of steel, aluminum, glass, or polyoxymethylene (POM) are employed.
Their diameters $D_s$ range from 20 mm to 40 mm, except for aluminum spheres, for which $D_s$ ranges from 15 mm to 30 mm.
The release height $H_s$ varies from 0.1 m to 1 m.
The density $\rho_s$ of each kind of sphere is determined by measuring the mass and the volume.
The contact angle $\theta_s$ between each material and water is measured using the static sessile drop method.
To further control surface wettability, selected spheres are coated with either Turtle Wax Super Hard Shell \citep{speirs2019water} or a silica nanoparticle coating, resulting in hydrophobic and hydrophilic wetting conditions, respectively.
The values of $\rho_s$ and $\theta_s$ are summarized in Table~\ref{tab:spheres}.

In addition to pure water, glycerol--water mixtures at three different volume fractions of glycerol are utilized as working liquids to vary the viscosity and surface tension.
The corresponding fluid properties, including viscosity $\mu$, surface tension coefficient $\sigma$, and density $\rho_l$ at around 20 $^\circ\text{C}$ are determined according to Ref.~\cite{Herran2010para}, and are listed in Table~\ref{tab:liquid parameters}.

\begin{table}
\caption{\label{tab:spheres}
Density $\rho_s$ and contact angle $\theta_s$ of spheres made of different materials.
The contact angle for spheres with either hydrophilic or hydrophobic coatings is also listed.
}
\begin{ruledtabular}
\begin{tabular}{lcc}
Material & $\rho_s\ (\mathrm{kg/m^3}$)& $\theta_s\ (^{\circ})$\\
\colrule
Steel                  & 7902     & $72\pm2$  \\
Aluminum              & 2710     & $83\pm3$  \\
Glass                  & 2253     & $30\pm1$  \\
Polyoxymethylene (POM) & 1355     & $92\pm2$  \\
Spheres coated with silica nanoparticle coating (Hydrophilic) & / & $21\pm3$ \\
Spheres coated with Turtle Wax Super Hard Shell (Hydrophobic) & / & $103\pm4$
\end{tabular}
\end{ruledtabular}
\end{table}

\begin{table}
\caption{\label{tab:liquid parameters}
Density $\rho_l$, surface tension coefficient $\sigma$, and viscosity $\mu$ of water and glycerol--water mixtures with different volume fractions.
}
\begin{ruledtabular}
\begin{tabular}{lccc}
Liquid & $\rho_l\ (\mathrm{kg/m^3}$)& $\sigma \ (\mathrm{mN/m})$& $\mu \ (\mathrm{mPa\ s})$ \\
\colrule
Water                 & 998     & 72.8 & 1.01 \\ 
10 vol\% Glycerol--water mixture (10\% G--W mixture)  & 1029   & 70.3 &1.26  \\
20 vol\% Glycerol--water mixture (20\% G--W mixture)  & 1060   & 69.1 &1.80  \\
30 vol\% Glycerol--water mixture (30\% G--W mixture)  & 1089   & 68.2 &2.71  \\
\end{tabular}
\end{ruledtabular}
\end{table}

\section{\label{sec:Qualitative analysis}Qualitative analysis}

\subsection{\label{subsec:Pinch-off modes}Pinch-off modes}

As the jet elongates rapidly in the vertical direction, three distinct modes emerge depending on $H_s$.
Fig.~\ref{Fig:snapshots}(a) shows a mode in which no child droplet is disengaged from the main liquid column throughout the entire process of rise and fall.
In contrast, a prominent child droplet detaches in the other two cases, indicating a pinch-off event.
Although the pinch-off phenomenon in the jet induced by sphere impact has been less extensively studied, insights can be found from the droplet impact problem.
For example, in droplet impact on a two-layer liquid system, two pinch-off modes were identified depending on the location at which the child droplet detaches \citep{kim2021impact}.
Inspired by this classification, when a child droplet separates during the falling stage, as shown in Fig.~\ref{Fig:snapshots}(b), the mode is termed ``downward pinch-off''.
In the opposite case (Fig.~\ref{Fig:snapshots}(c)), a child droplet has already formed before the jet reaches its maximum height, which is termed ``upward pinch-off''.
Notably, even after pinch-off occurs, the child droplet remains part of the jet and is taken into account in the analysis of jet height and shape.
The experimental parameters for Figs.~\ref{Fig:snapshots}(a--c) reveal a clear dependence of the pinch-off behavior on the release height, reflecting an intuitive physical mechanism: higher impact energies promote fluid destabilization.
According to the measured vertical velocity distribution reported by \citet{van2021self}, the upward velocity within the jet increases when approaching the jet tip, indicating a concentration of kinetic energy in the upper part of the jet.
When $H_s$ is relatively small, the impact transmits insufficient energy to the jet, implying that the upper part of the jet lacks the kinetic energy required to overcome surface tension and escape from the liquid column.
As $H_s$ increases, more kinetic energy is transferred from the sphere to the fluid, increasing the energy for jet formation.
After reaching maximum height, the lower part of the jet falls faster than the upper part due to its initially smaller upward velocity, as evidenced by the Supplemental Material~\citep{supplemental2025}.
Under the influence of surface tension, a neck forms between the two parts and subsequently pinches off, resulting in the formation of a child droplet and its separation from the main liquid column.
At even higher release heights, the jet gains a greater initial velocity, causing the upper part to ascend faster and separate from the lower part.

\begin{figure}
\centering
\includegraphics[width=460pt]{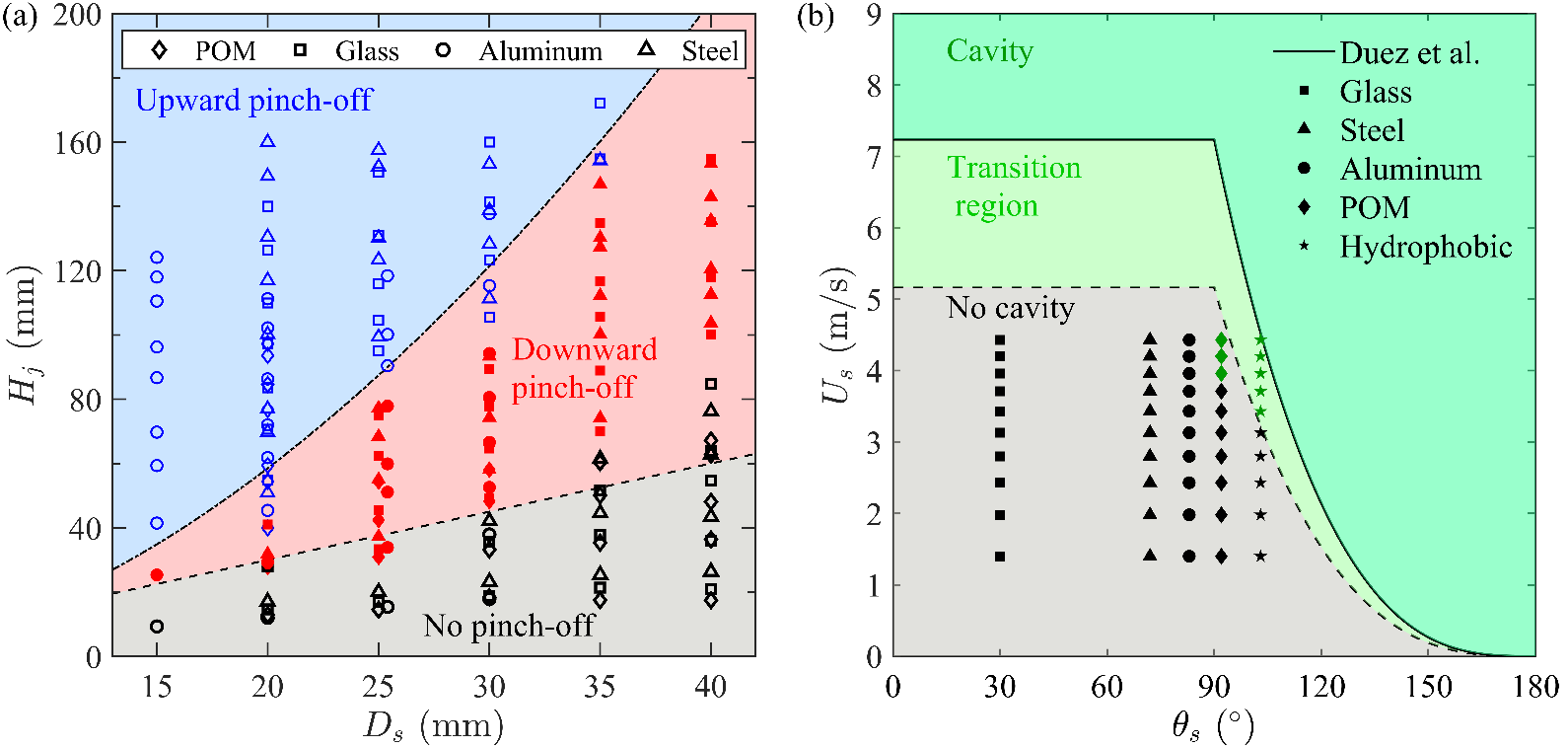}
\caption{\label{Fig:jet mode}(a) Regime diagrams illustrating the pinch-off modes of Worthington jets corresponding to four types of spheres.
Hollow black dots with thick edges: no pinch-off; solid red dots: pinch-off during the falling stage; hollow blue dots with thin edges: pinch-off during the rising stage.
Each dot corresponds to an averaged experimental data with at least 3 repeated tests.
The dashed and chain lines are transition boundaries $H_j^{cr1}$ and $H_j^{cr2}$, respectively, which are theoretically obtained from Eq.~(\ref{eq:critical Hj}).
(b) Impact velocity for cavity formation versus contact angle $\theta_s$.
Squares, triangles, circles, diamonds, and stars denote experimental results for spheres made of glass, steel, aluminum, POM, and hydrophobic coated spheres (any of the above materials with hydrophobic coating), respectively.
Diameter of the sphere is not shown explicitly.
The solid line corresponds to the theoretical curve given by \citet{duez2007making}.
The dashed boundary is obtained by adjusting the parameter $g_0$ in their mathematical model from 7 to 5 to fit the present experiments.
In both panels (a) and (b), water is used as the working liquid, with its corresponding fluid properties employed to obtain the theoretical curves.}
\end{figure}

To systematically characterize these pinch-off modes, experiments are conducted in wide ranges of $D_s$ and $H_s$ using spheres of four different materials, and the corresponding pinch-off positions are identified.
Since the subsequent theoretical analysis determines the critical value of the maximum jet height, rather than that of the release height, to distinguish between different pinch-off modes, we introduce the maximum jet height $H_j$, defined as the vertical distance between the undisturbed liquid surface and the jet tip when it reaches its highest point.
This is illustrated in the snapshot tagged ``112.5 ms'' in Fig.~\ref{Fig:snapshots}(b).
The experimental data of the maximum jet height $H_j$ against $D_s$ are plotted in Fig.~\ref{Fig:jet mode}(a), showing that the pinch-off behavior follows a similar regime pattern regardless of the material.
Because experimental evidence demonstrates that a higher release height leads to an increased maximum jet height, the preceding analysis relating $H_s$ to pinch-off modes remains applicable for $H_j$.
This supports the conjecture that pinch-off occurs only when $H_j$ exceeds a critical threshold, and the transition from downward to upward pinch-off takes place at a still higher $H_j$.
Additionally, the pinch-off exhibits a dependence on $D_s$: for larger spheres, a higher $H_j$ is required to achieve the mode transition, including both the trigger of pinch-off and the shift from downward to upward pinch-off.

Given that the pinch-off is governed by the competition between inertia and capillary forces, the Rayleigh--Plateau instability is applied.
This theory is supported by the experimental measurement that the diameter of the liquid column $D_j$ near the pinch-off point is approximately $3\pm1$ mm (Figs.~\ref{Fig:snapshots}(b--c)), which is comparable to the capillary length $l_c\equiv\sqrt{\sigma/\rho_l g}\approx2.7\ \mathrm{mm}$ for water.
According to Ref.~\cite{rayleigh1878instability}, the fastest-growing wavelength is approximated by $\lambda_m\approx4.5D_j$, and a local minimum in the interface profile can develop only when $H_j>\lambda_m/2$.
The diameter of the liquid column $D_j$ is estimated by assuming a cylindrical jet shape and expecting that its volume is approximately equal to that of the sphere, namely $\pi D_j^2 H_j/4\approx\pi D_s^3/6$.
Thus, $D_j$ is expressed in terms of $H_j$ such that $D_j\approx\sqrt{2D_s^3/3H_j}$, and a critical threshold $H_j^{cr1}$ is obtained by solving $H_j=9D_j/4$ for $H_j$, beyond which pinch-off occurs.
Above this threshold, the pinch-off time $T_p$ is approximated by the capillary-inertial timescale, i.e., $T_p\approx \sqrt{\rho_lD_j^3/\sigma}$, where $D_j$ has been taken as the characteristic length scale.
To clarify the critical boundary between the downward and upward pinch-off modes, we introduce the duration time from the initiation of the jet until it reaches its maximum height, which is denoted by $T_0$.
Because the evolution of the jet tip can be regarded as a uniformly decelerated motion under gravity (an assumption that will be validated subsequently), $T_0$ is calculated by $T_0=\sqrt{2H_j/g}$.
Different pinch-off modes are then determined by the relative magnitude of $T_p$ and $T_0$.
If $T_p>T_0$, the pinch-off occurs after the jet tip reaches its maximum height, corresponding to the downward pinch-off mode.
Conversely, the upward pinch-off mode appears when $T_p<T_0$.
The critical threshold $H_j^{cr2}$ between the two modes is obtained by solving $T_p=T_0$ for $H_j$.
The mathematical expressions of the two transition boundaries are summarized as follows:
\refstepcounter{equation}
\begin{equation}
H_j^{cr1}= \frac{3}{2} D_s, \qquad
H_j^{cr2}=\sqrt[5]{\frac{2\rho _l^2g^2D_s^9}{27\sigma ^2} }.
\tag{1a,1b} \label{eq:critical Hj}
\end{equation}

In Fig.~\ref{Fig:jet mode}(a), we have plotted the above theoretical curves, which partition the parameter space into three distinct regimes.
The experimental dots, classified by their observed pinch-off modes, are basically distributed within the corresponding theoretical regions.
Thus, the theoretical analysis based on the Rayleigh--Plateau instability provides a physical interpretation of the regime transitions observed in Fig.~\ref{Fig:jet mode}(a).

\subsection{\label{subsec:Critical condition for cavity formation}Critical condition for cavity formation}

Apart from the previously discussed pinch-off modes, the cavity may appear under specific experimental conditions, e.g., Fig.~\ref{Fig:snapshots}(d).
Different response behaviors of the liquid surface are observed at the initial stage of impact, visible in the first snapshot of each image sequence in Fig.~\ref{Fig:snapshots}.
In Figs.~\ref{Fig:snapshots}(a--c), the liquid climbs along the sphere surface and converges at the top, leading to the ejection of tiny droplets.
The liquid subsequently fills the wake region behind the sphere, and the transferred energy initiates jet formation.
In contrast, for the case shown in Fig.~\ref{Fig:snapshots}(d), the liquid separates from the sphere surface and forms a splash crown.
The air is then entrained into the liquid layer and fills the space vacated by the sphere, resulting in the development of a cavity.
If the cavity collapses beneath the liquid surface, a jet would be ejected from the pinch-off point \cite{gekle2010generation}, exhibiting a generation mechanism  fundamentally distinct from that observed in the present study.

The critical condition for cavity formation has been extensively investigated in previous studies \cite{Watson2018,speirs2018entry,watson2021water}, primarily based on the theoretical framework developed by \citet{duez2007making}.
The latter derived the critical impact velocity for air entrainment that depends on the contact angle of the impact body, which is plotted together with the current experimental data in Fig.~\ref{Fig:jet mode}(b).
For $\theta_s<90^{\circ}$, the critical velocity for cavity formation is identical and far beyond the impact velocity in our experiments, consistent with the observation that no cavity forms for spheres made of glass, steel, or aluminum.
However, when $\theta_s>90^{\circ}$, this critical velocity decreases nonlinearly with increased $\theta_s$, making cavity formation more likely for POM spheres.
Since spheres coated with Turtle Wax Super Hard Shell exhibit even larger $\theta_s$, the cavity can form at lower impact velocities.
Notably, hydrophobic spheres would create cavities at an impact velocity lower than the theoretical one.
By adjusting the parameter $g_0$ from 7 to 5 in the mathematical model proposed by \citet{duez2007making} to fit the present experiments, a modified boundary (dashed curve in Fig.~\ref{Fig:jet mode}(b)) between the cavity and cavity-free states is obtained.
A transition region, highlighted in light green in Fig.~\ref{Fig:jet mode}(b), is observed between this fitted boundary and the original theoretical curve obtained from \citet{duez2007making}.
This mixed region has been reported by \citet{truscott2009spin} and attributed to the effect of surface roughness.

\section{\label{sec:Jet dynamics}Jet dynamics}

\subsection{\label{subsec:Scaling analysis of maximum jet height}Scaling analysis of maximum jet height}

\begin{figure}
\centerline{\includegraphics[width=350pt]{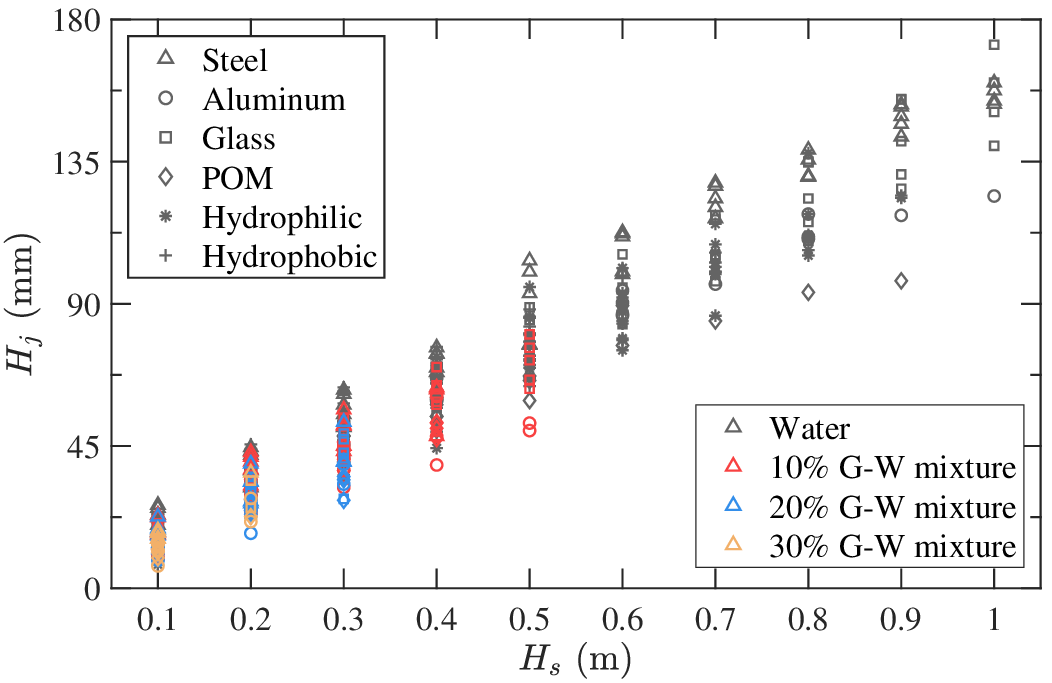}}
\caption{\label{Fig:Hj-Hs}Experimental relationship between the maximum jet height $H_j$ and the release height $H_s$.
The data include all experiments conducted with steel, aluminum, glass, POM, and coated spheres impacting both water and three different glycerol--water mixtures.}
\end{figure}

Having qualitatively characterized the geometric features of the jet, we now proceed to a quantitative investigation of its dynamics.
Based on the assumption that the kinetic energy of the sphere is converted into the potential energy of the jet, \citet{Watson2018} proposed a linear relation $H_j\propto H_s$, which, to our knowledge, has been the only scaling for this problem in the cavity-free regime.
In Fig.~\ref{Fig:Hj-Hs}, we plot this relationship for all our experimental data across various impact parameters and liquid properties, whose points are dispersed within a wide band.
Although the trend $H_j\propto H_s$ generally holds for each individual configuration, this simple linear relation neglects other governing parameters and therefore offers limited predictive accuracy.
Our goal here is to reveal the underlying physics and find a universal scaling that collapses all the data in Fig.~\ref{Fig:Hj-Hs}.

With the comprehensive set of experimental data, dimensional analysis provides a direct and effective method to relate the maximum jet height $H_j$ to the following experimental parameters: the density and diameter of the sphere $\rho_s$ and $D_s$, the impact velocity $U_s\approx \sqrt{2gH_s}$, gravitational acceleration $g$, contact angle of the sphere $\theta_s$, density $\rho_l$, surface tension coefficient $\sigma$, and viscosity $\mu$ of the working liquid.
By non-dimensionalizing with $D_s$, $\rho_l$, and $g$, the functional relationship between $H_j$ and the above parameters can be expressed as
\begin{equation}
\frac{H_j}{D_s} =f \left ( \tilde{\rho} , \mathrm{Fr}, \mathrm{We}, \mathrm{Re}, \theta_s  \right ),
\label{eq:dimensional relation}
\end{equation}
where $\tilde{\rho}=\rho_s/\rho_l$ is the density ratio, $\mathrm{Fr}=2 H_s/D_s$, $\mathrm{We}=2 g H_s \rho_l D_s/\sigma$, and $\mathrm{Re}=\sqrt{2gH_s} D_s/\nu$ are Froude, Weber, and Reynolds numbers, respectively, with $\nu=\mu/\rho_l$ denoting the kinematic viscosity of the liquid.
A comparison of the experimental results for spheres made of the same material under different coating conditions reveals that $H_j$ remains independent of $\theta_s$.
Consequently, the surface wettability primarily governs whether a cavity will form; once the jet generates without cavity formation, its subsequent dynamics show no discernible dependence on the contact angle.

\begin{figure}
\centering
\includegraphics[width=460pt]{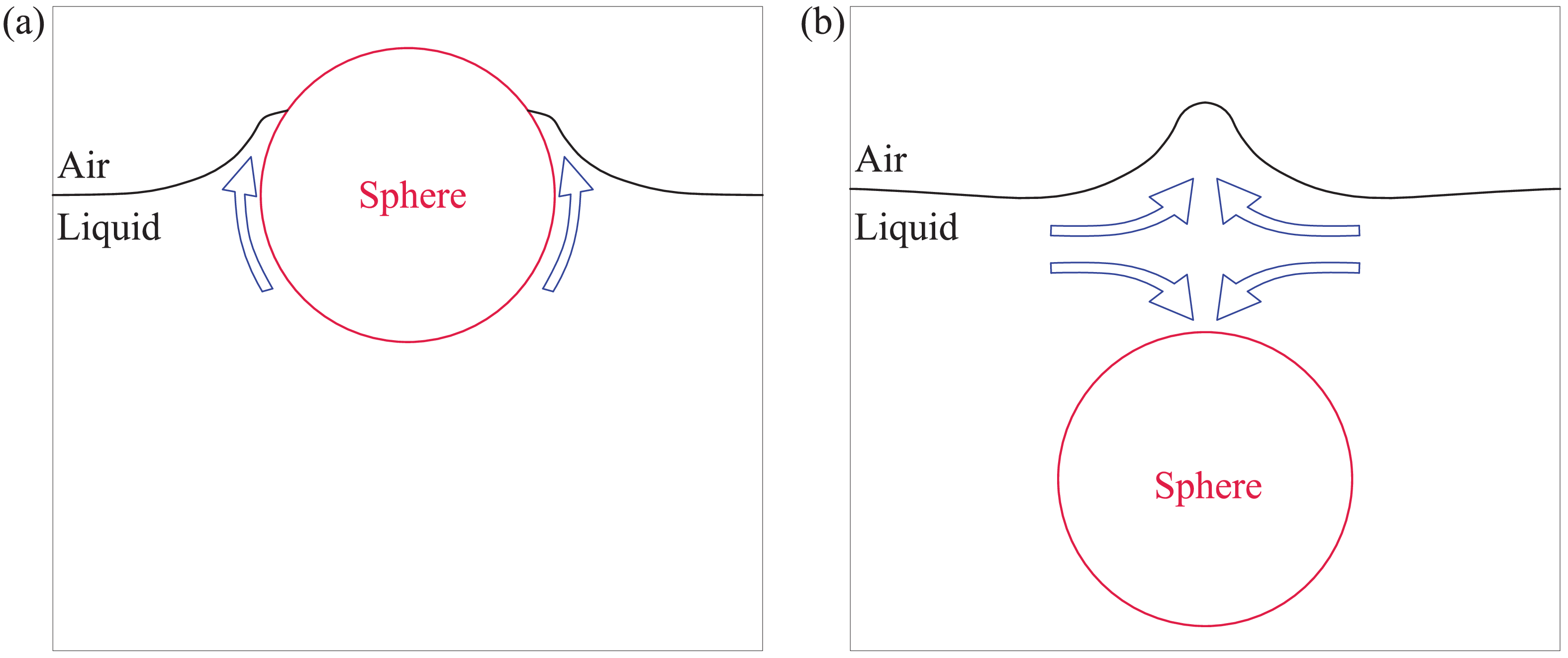}
\caption{\label{Fig:flow_field}Sketches of the flow near the free surface based on the numerical simulation by \citet{do2009splash}, illustrating (a) the initial impact stage when the sphere is partially immersed, and (b) the early jet stage after the sphere has moved away from the free surface.
Blue arrows indicate the flow direction.
}
\end{figure}

To derive a robust scaling law from Eq.~(\ref{eq:dimensional relation}), we present an analysis from the aspect of momentum and energy, beginning by clarifying the physical process of impact and jet formation.
In the cavity-free regime, \citet{do2009splash} provided the flow field obtained from numerical simulation, sketched in Fig.~\ref{Fig:flow_field} (referring to Fig.2 in Ref.~\cite{do2009splash} for full details).
During initial impact, when the sphere is not completely immersed (Fig.~\ref{Fig:flow_field}(a)), a thin liquid film climbs up the sphere surface and converges at the top, producing fine splashes and tiny droplets observed in the first two snapshots in Figs.~\ref{Fig:snapshots}(a--c).
Later, as the sphere recedes from the free surface (Fig.~\ref{Fig:flow_field}(b)), the liquid continuously accumulates toward the centerline behind the sphere and collides, generating upward and downward flow branches.
The upward branch evolves into the jet, while the downward branch forms a wake following the sphere.
This process highlights the essential role of the free surface in jet formation, since in an unbounded flow past a sphere the flow simply forms vortexes and decays downstream without producing such upward ejection.
For the two branches generated by the collision, the mass and characteristic velocity are accordingly of comparable orders, giving $m_w\sim m_j$ and $U_w\sim U_j$, where the subscripts $w$ and $j$ refer to the wake and the jet, respectively.
These relations imply an approximate momentum balance between the upward jet and the downward wake.
Just before and after impact, the momentum conservation reads:
\begin{equation}
m_sU_s=\left ( m_s+m_a \right ) U_f+m_wU_w-m_jU_j,
\label{eq:momentum}
\end{equation}
where the left-hand side is the momentum of the sphere before impact, with $m_s=\pi\rho_sD_s^3/6$ being its mass.
Upon immersion, whether the cavity forms or not, the momentum is partially transferred to the added mass $m_a=\pi C_m\rho_lD_s^3/6$ that moves at the same velocity $U_f$ as the sphere \cite{aristoff2009water}, where $C_m\approx0.5$ is the added mass coefficient \cite{newman1977marine}.
The second term $m_w U_w$ on the right-hand side of Eq.~(\ref{eq:momentum}) accounts for the momentum of the downward wake, an effect that has been incorporated into the force model by \citet{truscott2012unsteady} for the water entry problem of spheres.
The final term $-m_j U_j$ in Eq.~(\ref{eq:momentum}) represents the momentum of the upward jet.
Since the momentum of the upward jet is comparable to that of the downward wake, namely $m_jU_j \sim m_wU_w$, the velocity of the sphere together with the added mass is derived as $U_f \approx [\tilde{\rho}/(\tilde{\rho}+C_m)]U_s$, indicating that the velocity loss during impact decreases for spheres with a higher density.
To establish the relationship between $U_j$ and $U_s$, the energy conservation is considered as follows:
\begin{equation}
\frac{1}{2}m_sU_s^2 = \frac{1}{2}\left ( m_s+m_a \right ) U_f^2+ \frac{1}{2} m_wU_w^2 + \frac{1}{2} m_jU_j^2.
\label{eq:energy}
\end{equation}
Combining with the relationships $m_w \sim m_j\sim \pi\rho_lD_s^3/6$ and $U_w \sim U_j$, the initial velocity of the jet scales as $U_j^2 \sim [C_m \tilde{\rho}/(\tilde{\rho}+C_m)]U_s^2$.
Because the jet tip undergoes a uniformly decelerated motion under gravity, $U_j$ satisfies $U_j\approx \sqrt{2g H_j}$.
Hence, a scaling for the maximum jet height is obtained, which reads
\begin{equation}
H_j \propto \frac{\tilde{\rho}}{\tilde{\rho}+C_m} H_s.
\label{eq:Hj-Hs}
\end{equation}

Next, we rewrite Eq.~(\ref{eq:dimensional relation}) in the form $H_j/D_s = K [\tilde{\rho}/(\tilde{\rho}+C_m)] \mathrm{Fr}^a \mathrm{We}^b \mathrm{Re}^c$ (with $K$ a proportionality constant) and expand it as $H_j\propto [\tilde{\rho}/(\tilde{\rho}+C_m)] H_s^{a+b+c/2 } D_s^{-a+b+c+1} \sigma^{-b} \nu^{-c}$.
Equation~(\ref{eq:Hj-Hs}) indicates that $H_j$ is proportional to $H_s$ and independent of $D_s$, resulting in the constraints $a+b+c/2 =1$ and $-a+b+c+1= 0$.
However, these two conditions alone are insufficient to uniquely determine all three exponents.
Due to the limited theoretical foundation for cavity-free Worthington jets, we seek guidance from analogous phenomena.
\citet{thoroddsen2004impact} investigated the horizontal jet ejected radially immediately following the initial impact of the sphere, whose experimental results report that the normalized initial velocity of this jet scales as $U_{jet}/U_{sphere} \propto \mathrm{Re} ^ \alpha$ with $\alpha \approx 0.245$.
Although the jet geometry differs substantially from that in the present problem, this scaling suggests an available functional dependence of $H_j/D_s$ on $\mathrm{Re}$.
Recalling the velocity-height relations $U_j\approx \sqrt{2gH_j}$ and $U_s\approx \sqrt{2gH_s}$, we adopt the scaling $H_j \propto \mathrm{Re}^{0.5} H_s$.
Combining this with the previous constraints yields the exponents $a=1.125$, $b=-0.375$, and $c=0.5$.
The final scaling law for the dimensionless maximum jet height is written as
\begin{equation}
\frac{H_j}{D_s}= K \frac{\tilde{\rho}}{\tilde{\rho}+C_m} \mathrm{Fr}^{1.125} \mathrm{We}^{-0.375} \mathrm{Re}^{0.5}.
\label{eq:scaling law}
\end{equation}

\begin{figure}
\centering
\includegraphics[width=350pt]{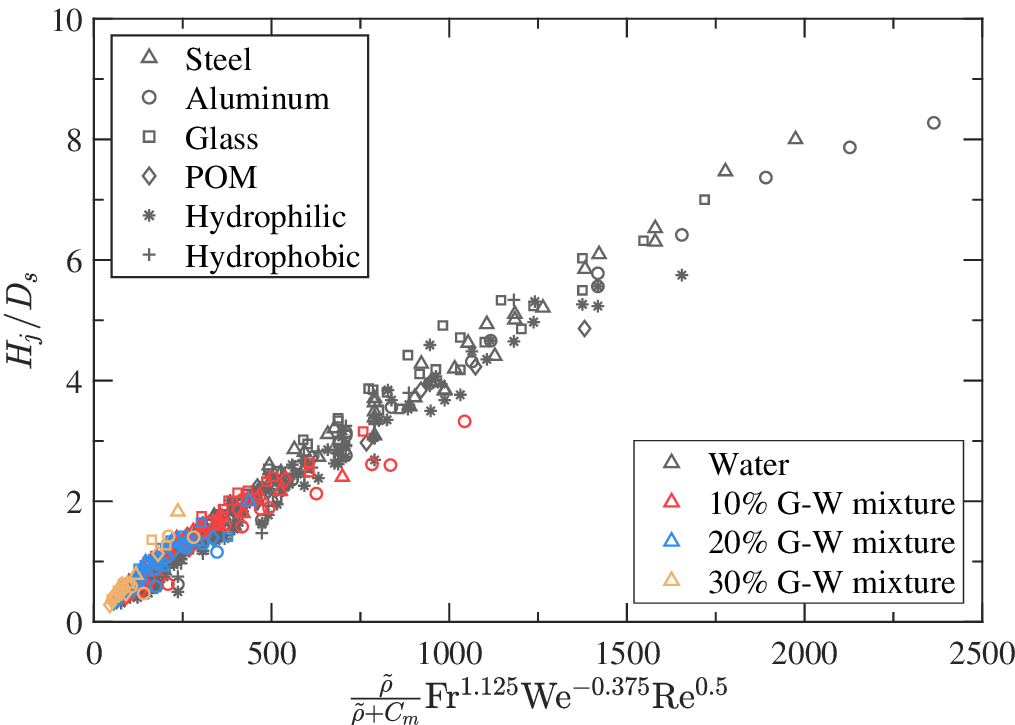}
\caption{\label{Fig:scaling}Experimental relationship between the dimensionless maximum jet height $H_j/D_s$ and the combined dimensionless parameters defined in Eq.~(\ref{eq:scaling law}).
The data include all experimental configurations as described in Fig.~\ref{Fig:Hj-Hs}.}
\end{figure}

It should be emphasized that Eq.~(\ref{eq:scaling law}) is a theoretical result derived from first principles, with no experimental data employed in advance to determine the exponents.
Therefore, this scaling law is entirely a priori, and the experimental results serve solely for its validation.
Figure~\ref{Fig:scaling} presents the experimental relationship between $H_j/D_s$ and the combined dimensionless parameters defined in Eq.~(\ref{eq:scaling law}), which spans all series of experiments conducted with spheres of different materials and liquids of varying properties.
The experimental data from these diverse conditions collapse remarkably well onto a single master curve, thus confirming the validity and universality of the proposed scaling law.
Although the linear relationship $H_j\propto H_s$ is retained in Eq.~(\ref{eq:scaling law}), it explicitly incorporates the additional parameters that determine the jet height, significantly improving predictive precision compared to the results in Fig.~\ref{Fig:Hj-Hs}.
Equation~(\ref{eq:scaling law}) indicates that the maximum jet height is governed primarily by the Froude number, highlighting the dominant role of jet's inertia competing against gravity in the jet dynamics.
Viscous and surface tension effects also play important roles.
Viscosity acts as a dissipation that suppresses jet formation and development \citep{C9SM00318E,10.1063/5.0084456}.
During the initial impact, surface tension acts immediately on the highly curved liquid surface (i.e. the formation of both a rapid local bulge and a transient circular crater around the sphere, as shown in Fig.~\ref{Fig:snapshots}(a--c) and Fig.~\ref{Fig:flow_field}(a)).
A higher surface tension coefficient corresponds to greater surface energy that the fluid carries, enabling more impact energy to be absorbed by the fluid and accumulated in the free surface region to feed the upward jet.

\subsection{\label{subsec:Evolution of jet height and shape}Evolution of jet height and shape}

The scaling analysis clarifies the parametric dependence of the maximum jet height, but provides no information on the temporal evolution.
We therefore proceed to a more detailed investigation of the time-dependent development of the jet height and shape.
In the context of the entry problem of a solid object, most research has focused on cavity dynamics, which was studied theoretically by solving the Rayleigh--Besant problem \cite{duclaux2007dynamics,aristoff2009water,zou2024cavity} or numerically using boundary integral simulations \cite{gekle2009high,gekle2010generation}.
Although such simulations provide valuable insight into jet formation driven by cavity collapse, they are not directly applicable to the cavity-free jet studied here.
Examining the jet profiles shown in Figs.~\ref{Fig:snapshots}(a--c), we observe that the jet maintains a slender and predominantly vertical structure and exhibits a basically self-similar shape, particularly during the rising phase.
This behavior resembles that of the jet generated by droplet impact on a liquid surface \cite{kim2021impact,michon2017jet,che2018impact, van2018emanating, shahdhaar2025low}, where self-similar jet profiles have been theoretically analyzed.
Building on these foundations, we present an analysis of the jet evolution.

For such axisymmetric liquid columns, cylindrical coordinates are utilized, where $z$ is set along the upward direction, with $z=0$ representing the undisturbed free surface.
The one-dimensional governing equations for momentum and mass conservation that have incorporated both surface tension and gravity read \cite{eggers1994drop}
\begin{subequations}
\label{eq:momentum&mass}
\begin{equation}
\frac{\partial u}{\partial t} +u\frac{\partial u}{\partial z} =-\frac{1}{\mathrm{Bo}} \frac{\partial \kappa }{\partial z}-1,
\label{eq:momentum_evolution}
\end{equation}
\begin{equation}
\frac{\partial r}{\partial t} +u\frac{\partial r}{\partial z} =-\frac{1}{2}r \frac{\partial u }{\partial z},
\label{eq:mass}
\end{equation}
\end{subequations}
where $u(z,t)$ denotes the local axial velocity within a liquid slice, $r(z,t)$ is the jet radius, and $\kappa(z,t)\equiv 1/r(z,t)$ represents the leading order curvature.
All variables have been non-dimensionalized using $D_s$, $\rho_l$, and $g$, with the scaled quantities: $u=u'/\sqrt{gD_s}$, $r=r'/D_s$, $\kappa=\kappa'D_s$, $z=z'/D_s$, and $t=t'\sqrt{g/D_s}$.
The Bond number in Eq.~(\ref{eq:momentum_evolution}) is defined as $\mathrm{Bo}=\rho_l g D_s^2/\sigma$.
Notably, the viscous effects, which are accounted for in Eq.~(\ref{eq:scaling law}), are not retained in Eqs.~(\ref{eq:momentum&mass}).
This difference does not represent a conflict but rather reflects the distinct importance of viscosity at the different physical stages addressed by each theory.
The scaling law in Eq.~(\ref{eq:scaling law}) focuses on the influence of impact dynamics on the maximum jet height.
In this impact stage, the viscosity is non-negligible to model the flow around a sphere.
The Reynolds number is thus retained to incorporate the experimental fact that higher viscosity of the working liquid results in a lower maximum jet height under identical impact conditions, confirming that viscosity is non-negligible during the impact stage.
In contrast, Eqs.~(\ref{eq:momentum&mass}) serve to describe the jet evolution after its initial formation.
As will be demonstrated subsequently, this evolution is primarily governed by gravity, which justifies the exclusion of viscous effects.
Therefore, viscosity plays a pronounced role only during the impact stage; once the jet begins to rise, the influence of viscosity on its subsequent motion becomes negligible.

Equations~(\ref{eq:momentum&mass}) are nonlinear partial differential equations that cannot be directly solved for analytical solutions.
Following the approach developed by \citet{ting1990slender} and \citet{van2021self}, we construct self-similar solutions based on scale-invariance arguments.
One set of particular solutions for Eqs.~(\ref{eq:momentum&mass}) is written as (see Appendix~\ref{appendix} for a detailed derivation):
\begin{subequations}
\label{eq:solutions}
\begin{equation}
u\left ( z,t \right ) =\frac{z}{t} -\frac{t}{2} +B\sqrt{t},
\label{eq:u}
\end{equation}
\begin{equation}
r\left ( z,t \right ) =\frac{4\sqrt{t} }{3B \mathrm{Bo} \left ( 4Bt^{3/2}-t^2-2z\right )},
\label{eq:r}
\end{equation}
\begin{equation}
\kappa \left ( z,t \right ) =-3B\mathrm{Bo}\left ( \frac{z}{2\sqrt{t} } +\frac{t^{3/2}}{4} -Bt \right ).
\label{eq:kappa}
\end{equation}
\end{subequations}

\begin{figure}
\centering
\includegraphics[width=460pt]{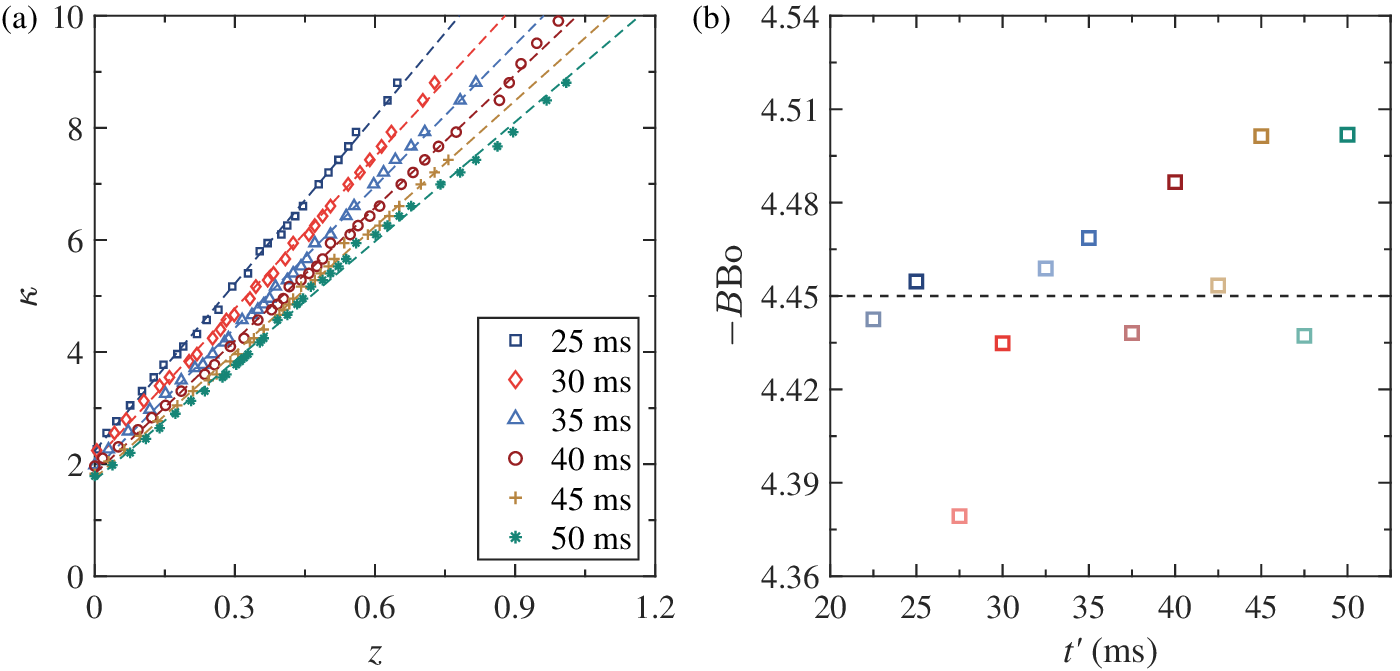}
\caption{\label{Fig:determine_B}(a) Experimental results of the dimensionless jet curvature $\kappa$ as a function of vertical location $z$ at different times, with data sets spaced at a 5 ms interval.
Dashed lines represent linear fits to corresponding data.
The experimental configuration is a steel sphere of $D_s=30\ \mathrm{mm}$ impacting the water surface at $H_s=0.5\ \mathrm{m}$.
(b) Values of $-B\mathrm{Bo}$ determined from the slopes of $\kappa$-$z$ curves at different times.
The horizontal dashed line at $-B\mathrm{Bo}=4.45$ indicates approximately the average of all values.
}
\end{figure}

The coefficient $B$ in Eqs.~(\ref{eq:solutions}) remains undetermined by the current theoretical constraints.
To establish an analogous parameter, \citet{van2021self} employed experimentally measured velocity distributions and fitted them to a theoretical expression in the form of Eq.~(\ref{eq:u}).
Similarly, the model developed by \citet{ghabache2014liquid} introduced a constant prefactor determined by averaging the fitted values of several different experimental configurations.
Inspired by these approaches, we determine $B$ by comparing with experimental data.
Equation~(\ref{eq:kappa}) shows a simple functional relationship between $\kappa$ and $z$: at any given time, $\kappa$ varies linearly with $z$, having a slope of $-3B \mathrm{Bo} /2\sqrt{t} $.
This explicit dependence indicates that the slope of the experimental $\kappa$-$z$ curve provides a direct measurement of $B$.
Selecting the experimental data for a steel sphere of $D_s=30\ \mathrm{mm}$ impacting the water surface at $H_s=0.5\ \mathrm{m}$, we extract jet profiles between 22.5 ms and 50 ms and calculate $\kappa$ for varying $z$.
Several series of experimental data are plotted in Fig.~\ref{Fig:determine_B}(a), which align closely with their respective linear fits.
The slopes of the fitted lines are then used to calculate the values of $-B\mathrm{Bo}$, which are summarized in Fig.~\ref{Fig:determine_B}(b).
The results obtained at different times fluctuate around a mean of 4.45 with a deviation of $\pm0.07$, leading to the estimation $B\approx-4.45/\mathrm{Bo}$.
Substituting this expression into Eqs.~(\ref{eq:solutions}) gives the final unique form of solutions.

Although self-similar solutions effectively describe the jet shape along the $z$-direction, their limitations become apparent when examining the jet height.
Equation~(\ref{eq:r}) predicts a hyperbolic jet profile whose radius asymptotically approaches zero as $z\to \infty$, resulting in the absence of a well-defined jet tip.
This formulation inherently loses information about the axial extension of the jet and cannot clarify the temporal evolution of the jet height.
Previous studies relying on similar assumptions have therefore been restricted to comparisons of the jet flanks rather than complete spatial characterization \cite{ghabache2014liquid,van2021self}.
To address this limitation, we introduce a kinematic condition to construct a relation between the flow velocity and the temporal evolution of the jet height.
At the jet tip, the vertical fluid velocity must be equal to the upward propagation speed of the free surface, implying $u\left ( h,t \right ) \equiv \dot{h } \left ( t \right )$, where $u\left ( h,t \right )$ is the fluid velocity at $z=h(t)$, with $h(t)$ being the instantaneous position of the jet tip that has been non-dimensionalized with $D_s$ (i.e., $h(t)=h'(t)/D_s$), and $\dot{h } \left ( t \right )$ denotes its time derivative.
Substituting Eq.~(\ref{eq:u}) into this condition, an ordinary differential equation for $h(t)$ is obtained, which is solved to yield
\begin{equation}
h\left ( t \right ) = U_0 t-\frac{1}{2}t^2-\frac{8.9}{\mathrm{Bo}}t^{3/2},
\label{eq:h_t}
\end{equation}
where the first term is associated with the dimensionless initial velocity of the jet $U_0=U_0'/\sqrt{gD_s}$ through the condition $\dot{h } \left ( 0 \right )=U_0$, while the second and third terms account for the effects of gravity and surface tension, respectively.
The determination of $U_0$ depends critically on whether the surface tension is included in Eq.~(\ref{eq:h_t}).
When surface tension is neglected, Eq.~(\ref{eq:h_t}) reduces to a uniformly decelerated motion under gravity.
To ensure that the maximum of $h'(t)$ is equal to the measured $H_j$, we have $U_0' = \sqrt{2gH_j}$, which coincides with the previous estimate $U_j\approx \sqrt{2g H_j}$.
When surface tension is incorporated, its restoring action restrains the upward motion of the jet.
If $U_0'=\sqrt{2gH_j}$ is retained, the predicted maximum of $h'(t)$ would fall short of $H_j$.
To resolve this phenomenon, we introduce a revised initial velocity $U_0'=C\sqrt{2gH_j}$ with $C>1$ being a numerical correction factor, whose value is determined to be $C\approx1.05$ by fitting with an arbitrary experimental configuration.

\begin{figure}
\centering
\includegraphics[width=460pt]{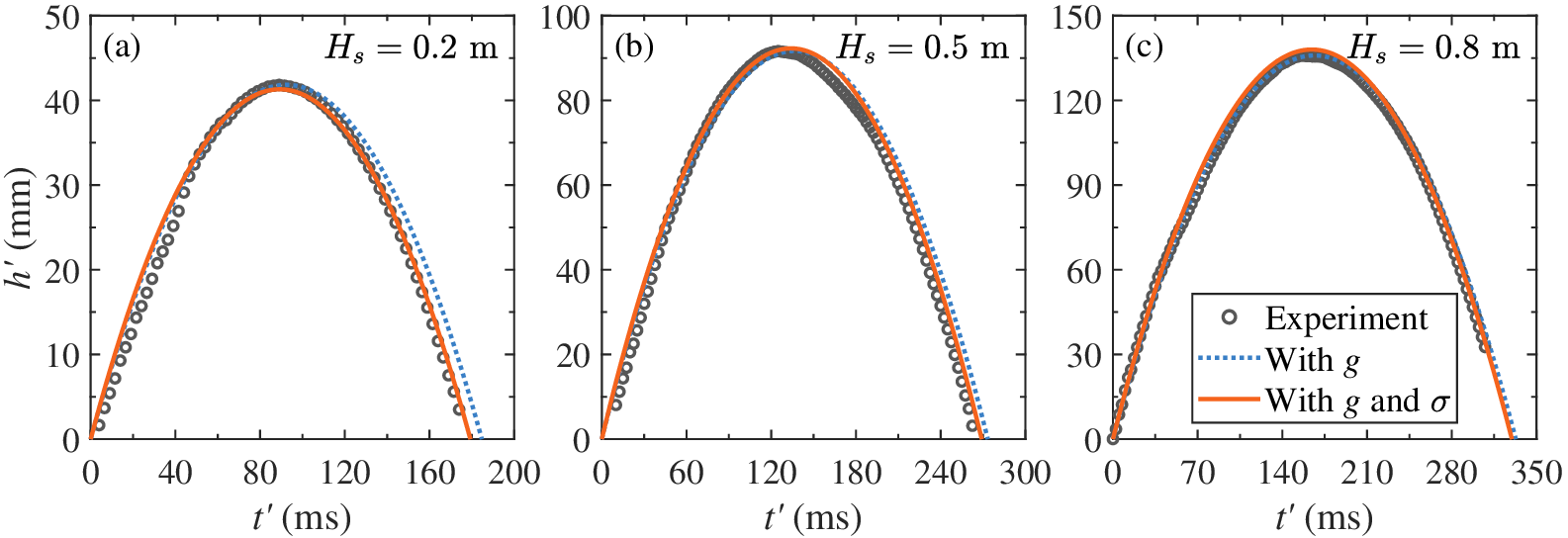}
\caption{\label{Fig:h_t}Comparison of the evolution of the dimensional jet height $h'(t)$ obtained from Eq.~(\ref{eq:h_t}) with experimental measurements.
Solid lines represent predictions incorporating both gravity and surface tension, while dotted lines show results considering gravity alone.
Experimental data presented by circles correspond to a steel sphere of $D_s=30\ \mathrm{mm}$ impacting the water surface at different release heights: (a) $H_s=0.2\ \mathrm{m}$, (b) $H_s=0.5\ \mathrm{m}$, and (c) $H_s=0.8\ \mathrm{m}$.
}
\end{figure}

\begin{figure}
\centering
\includegraphics[width=460pt]{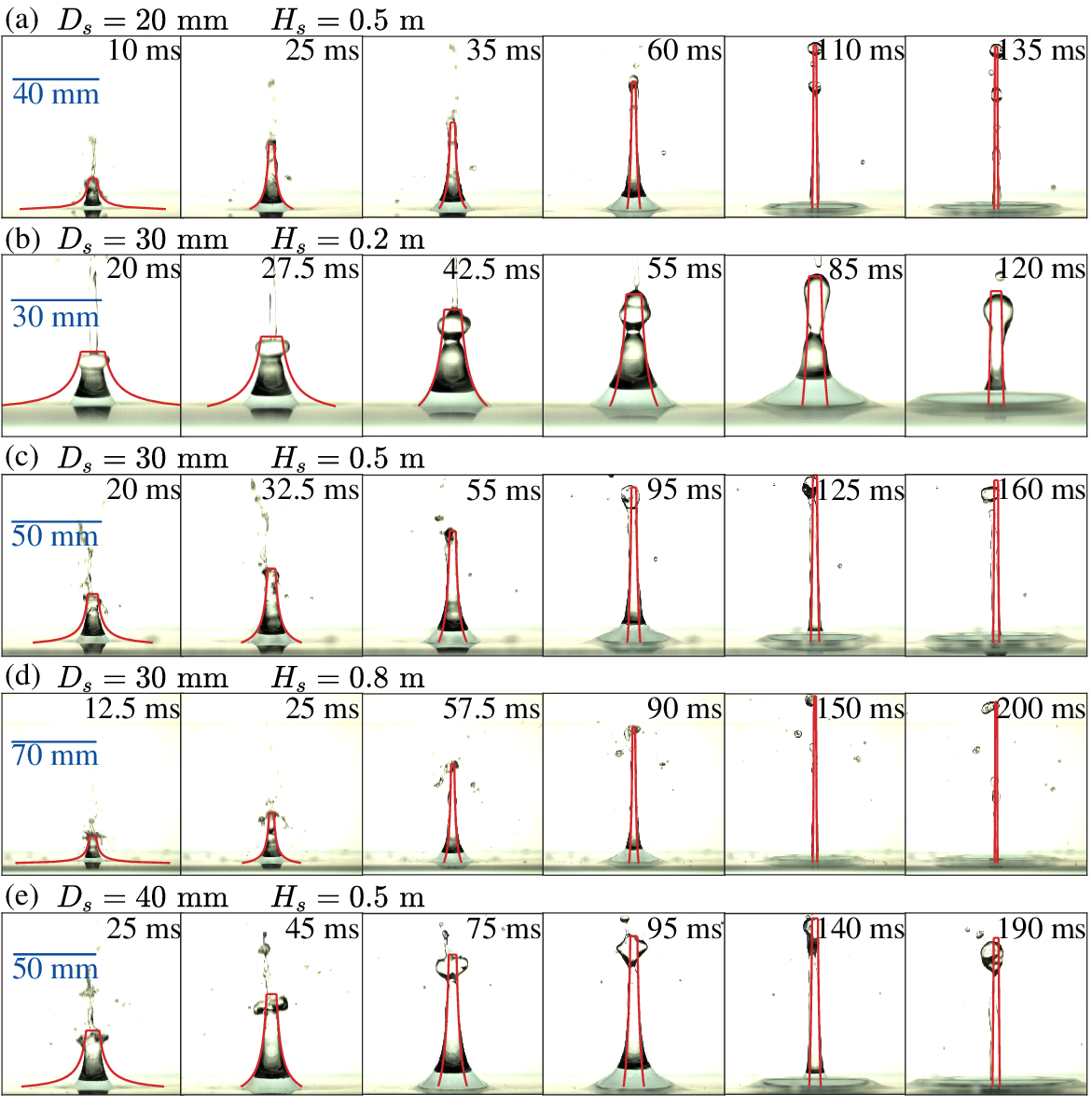}
\caption{\label{Fig:jet_shape}Comparison of theoretically predicted jet shapes (solid lines) with corresponding experimental snapshots.
The experimental configurations are steel spheres impacting the water surface, with sphere diameter $D_s$ and release height $H_s$ being reported in the figure.
}
\end{figure}

A comparison between theoretical $h'(t)$ and experiments is presented in Fig.~\ref{Fig:h_t}, which contains results for a steel sphere of $D_s=30\ \mathrm{mm}$ impacting the water surface at three different release heights.
Overall, the evolution of the jet height is reasonably well described by the uniformly decelerated motion under gravity, particularly at higher release heights.
This observation supports previous treatments of the characteristic time $T_0$ in Section~\ref{subsec:Pinch-off modes} and the jet velocity $U_j$ in Section~\ref{subsec:Scaling analysis of maximum jet height}.
Depending on the release height, surface tension provides a significant correction to the jet height, which is more obvious during the falling stage.
At low release heights where no pinch-off occurs (e.g., Fig.~\ref{Fig:snapshots}(a)), the continuous liquid column experiences a downward restoring force due to surface tension, accelerating the descent of the jet tip, as evidenced in Fig.~\ref{Fig:h_t}(a) by the substantial deviation from the gravity-only model.
As the release height increases sufficiently to trigger downward pinch-off (e.g., Fig.~\ref{Fig:snapshots}(b)), the restoring action of surface tension no longer acts directly on the jet tip defined by the top of the child droplet, leading to a weaker modification, as seen in Fig.~\ref{Fig:h_t}(b).
When the release height exceeds the critical threshold for the upward pinch-off (e.g., Fig.~\ref{Fig:snapshots}(c)), an earlier detachment helps the child droplet escape from the restoring action of surface tension, thus yielding the improved agreement between the experimental data and the theoretical model with gravity alone, as shown in Fig.~\ref{Fig:h_t}(c).
The difference between the models with and without surface tension decreases slightly in Fig.~\ref{Fig:h_t}(b) and significantly in Fig.~\ref{Fig:h_t}(c).
This is attributed to the fact that the detachment usually occurs at higher release heights (or maximum jet heights, see Fig.~\ref{Fig:jet mode}(a)), which leads the theory to predict a more slender jet shape, as shown in Figs.~\ref{Fig:jet_shape}(c--d).
In this regime, the term $\partial_z \kappa$ in Eq.~(\ref{eq:momentum_evolution}) associated with surface tension becomes negligible, which explains the closer consistence between the theories.
In summary, surface tension generally suppresses the rising motion of the jet and accelerates its descent, with the influence being most pronounced in the absence of pinch-off.
Once a child droplet detaches from the continuous main liquid column, gravity becomes the dominant force governing its subsequent motion.
Furthermore, we note some discrepancies between theory and experiment in Figs.~\ref{Fig:h_t}(a--b): the theoretical $h'$ rises faster than the experimental one in (a) and falls more slowly in (b).
This difference is attributed to the complex, unstable deformation of the child droplet, which remains either connected with the lower main liquid column in case (a) or detached in (b).
The first four snapshots in Fig.~\ref{Fig:jet_shape}(b) show that the ``droplet'' at the jet top (though not detached) evolves from a flatter to a slender shape, while the last three snapshots in Fig.~\ref{Fig:jet_shape}(c) reveal the separated droplet transitioning from slender to flatter.
These observations suggest a combined action of gravity, surface tension, and internal flow, probably producing a downward net force on the jet tip that exceeds gravity alone, which is not accounted for in the theory.

Combining the theoretical expressions for the jet radius $r(z,t)$ from Eq.~(\ref{eq:r}) and the jet height $h(t)$ from Eq.~(\ref{eq:h_t}), we now assess the model's ability to describe the evolution of jet shape.
At a given instant $t_0$, the jet height $h(t_0)$ is first evaluated from Eq.~(\ref{eq:h_t}); the jet profile $r(z,t_0)$ over $z\in  [0,h(t_0)]$ is then calculated from Eq.~(\ref{eq:r}).
Figure~\ref{Fig:jet_shape} compares these predicted jet profiles with the corresponding experimental snapshots at different times for steel spheres of varying diameters impacting the water surface at different release heights.
It should be noted that the self-similar theory assumes a continuous liquid column that includes the detached child droplet, and hence cannot capture the pinch-off event.
Overall, the theoretical results agree well with experiments across the tested parameter range.
For a fixed $D_s$ (e.g., Figs.~\ref{Fig:jet_shape}(b--d)), the parameters in Eq.~(\ref{eq:r}) remain unchanged, thus producing identical theoretical shapes but different height evolutions.
At a fixed $H_s$ (e.g., Figs.~\ref{Fig:jet_shape}(a,c,d)), the theory captures the dependence of jet shape on diameter: the jet becomes more slender for smaller $D_s$ and thicker as $D_s$ increases.
Both trends with varying $D_s$ and $H_s$ are well reproduced, which supports the self-similar framework.
Temporally, the theory predicts the upper portion of the jet more accurately than the lower part.
The experimental observations (e.g., Fig.~\ref{Fig:jet_shape} and Supplemental Material~\citep{supplemental2025}) suggest that as the jet evolves, a pronounced wave swell emerges on the liquid surface near the jet base and propagates outward from the impact point, which plays a key role in the observed deformation.
%Because both gravity and surface tension included in the model act as restoring forces, the theory tends to predict a faster descent of the liquid near the base than observations.
This discrepancy likely points to another physical effect: the impact process continuously supplies momentum to the liquid, resulting in a substantial deformation of the liquid surface near the jet base and the propagation of wave swell.
In contrast, the upper region elongates into a slender, nearly vertical profile where gravity dominates over surface tension, leading to a simpler shape that is well captured by the theory.
This clarifies the domain of applicability of the self-similar formulation.
While the self-similar theory has been widely used to model slender jets generated from droplet impact on a liquid surface \cite{van2021self}, its individual focus on the jet itself neglects the influence of impact process.
In the sphere impact circumstance, the flow behind the sphere continuously converges and feeds energy into the jet during its early stage, a mechanism not accounted for in the theory.
Bridging the self-similar model with the impact dynamics, particularly near the jet base, represents a promising direction for research.
Finally, the distinct generation mechanisms between cavity-free jets and those driven by cavity collapse are highlighted again.
In the absence of a cavity, the jet is significantly affected by the sphere and its trailing flow, fundamentally differing from the jet generated by cavity implosion as reported in \cite{gekle2009high,gekle2010generation}.

\section{\label{sec:Conclusions}Conclusions}

Compared with Worthington jets driven by cavity collapse, the jets generated during solid sphere impact on a liquid surface with no cavity formed have received relatively little attention.
In this study, we have validated that such jets arise from a fundamentally different mechanism.
Rather than being powered by the implosion of a cavity dominated mainly by surface tension, these jets are sustained by the collision of flow behind the sphere.
Upon impact, the kinetic energy of the sphere is partially lost to accelerate the added mass, and another portion concentrates liquid toward the centerline behind the sphere.
The collision of this converging flow produces an upward jet and a downward wake following the sphere, whose momenta are approximately balanced.
Once formed, the jet evolution is governed primarily by gravity and modified by surface tension, whose interplay can result in the detachment of a child droplet from the main liquid column, indicating a pinch-off event.

A qualitative analysis of the experimental results identifies three distinct pinch-off modes, which are independent of the sphere wettability and density.
Based on the Rayleigh--Plateau instability, the critical boundaries between these modes are theoretically determined by introducing the fastest-growing wavelength and estimating the capillary-inertial pinch-off timescale.
To clarify the underlying jet dynamics, a scaling analysis is performed grounded in momentum and energy conservation and is combined with existing experimental findings, eventually leading to a scaling law for the maximum jet height: $H_j/D_s= K [\tilde{\rho}/(\tilde{\rho}+C_m)] \mathrm{Fr}^{1.125} \mathrm{We}^{-0.375} \mathrm{Re}^{0.5}$.
All of the experimental data covering different kinds of spheres, release heights, and working liquids converge remarkably well to this predicted relationship.
This scaling confirms the dominance of inertial force in jet dynamics, while the fluid viscosity acts as a dissipation to suppress jet formation, and surface tension facilitates energy accumulation at the free surface.
Finally, self-similar solutions based on scale-invariance arguments are combined with the kinematic condition at the jet tip to predict the evolution of jet height and shape, which agree well with experiments.
The jet height is governed primarily by gravity, while surface tension provides a downward restoring force.
The influence of surface tension is most pronounced in the absence of pinch-off; otherwise, the detachment of a child droplet from the main liquid column renders its subsequent motion completely gravity dependent.
A limitation of this theory becomes apparent when comparing the jet shape: the model focuses on the jet itself and does not fully incorporate the ongoing momentum input from the impact process during the early stage.
Addressing this coupling represents a promising direction for research.
These insights may advance the fundamental understanding of jet driven by impact and be beneficial to applications in industrial, environmental, and biomechanical contexts.

\begin{acknowledgments}

The authors acknowledge the preliminary work of Yufei Zhang, Xue Jiang, and Mengyuan Jin.
This work was supported by the Fundamental Research Funds for the Central Universities.

\end{acknowledgments}

\appendix

\section{\label{appendix}Self-similar solutions based on scale-invariance arguments}

In Section~\ref{subsec:Evolution of jet height and shape}, Eqs.~(\ref{eq:momentum&mass}) governing the jet dynamics are solved analytically based on scale-invariance arguments, following the methodology of \citet{ting1990slender} and \citet{van2021self}.

We first consider the case where only surface tension is retained, reducing Eqs.~(\ref{eq:momentum&mass}) to
\refstepcounter{equation}
\begin{equation}
\frac{\partial u_1}{\partial t} +u_1\frac{\partial u_1}{\partial z} =-\frac{1}{\mathrm{Bo}} \frac{\partial \kappa_1 }{\partial z},\qquad 
\frac{\partial \kappa^{-1}_1}{\partial t} +u_1\frac{\partial \kappa^{-1}_1}{\partial z} =-\frac{1}{2}\kappa^{-1}_1 \frac{\partial u_1 }{\partial z}.
\tag{A1a,A1b} \label{appen:eq:no gravity}
\end{equation}
Introducing the scaling transformations $u_1\to u_1^*\bar{u}_1$, $\kappa_1 \to \kappa_1^*\bar{\kappa}_1$, $z\to z^*\bar{z}$, and $t\to t^*\bar{t}$, and imposing the scaling invariance of Eqs.~(\ref{appen:eq:no gravity}), we obtain the relations $u_1^*=z^*/t^*$ and $\kappa_1^*=(z^*/t^*)^2$.
The functional relation $(u_1,\kappa_1)=F(z,t)$ therefore becomes
\begin{equation}
\left(\frac{z^*}{t^*} \bar{u}_1 ,\left ( \frac{z^*}{t^*} \right )^2 \bar{\kappa}_1 \right)=F(z^*\bar{z},t^*\bar{t}),
\label{appen:eq:function_F}
\end{equation}
which holds for arbitrary choices of scales $z^*$ and $t^*$.
According to \citet{ting1990slender}, the relationship $z^*=t^{*\beta}$ is imposed with $\beta=3/2$.
Substituting this into Eq.~(\ref{appen:eq:function_F}) and setting $\bar{t}= 1/t^*$ gives
\begin{equation}
\left(\frac{\bar{u}_1}{\sqrt{\bar{t}}}  ,\frac{\bar{\kappa}_1}{\bar{t}}  \right) = G(\frac{\bar{z}}{\bar{t}^{3/2}}).
\label{appen:eq:function_G}
\end{equation}
Choosing a particular family of solutions, we have $u_1(z,t)=\sqrt{t}f_1(\xi)$ and $\kappa_1(z,t)=t g_1(\xi)$, with the self-similar variable $\xi=z/t^{3/2}$.
Assuming linear forms $f_1(\xi)=A\xi+B$ and $g_1(\xi)=C\xi+D$ and substituting them into Eqs.~(\ref{appen:eq:no gravity}), the coefficients are determined as $A=1$, $C=-3B\mathrm{Bo}/2$, and $D=3B^2\mathrm{Bo}$.
The solutions of Eqs.~(\ref{appen:eq:no gravity}) are summarized as:
\refstepcounter{equation}
\begin{equation}
u_1\left ( z,t \right ) =\frac{z}{t} +B\sqrt{t},\qquad 
\kappa_1 \left ( z,t \right ) = -\frac{3B\mathrm{Bo}z}{2\sqrt{t} } +3 B^2\mathrm{Bo} t .
\tag{A4a,A4b} \label{appen:eq:solutions_1}
\end{equation}

We now reintroduce the gravity term and seek solutions to the complete system Eqs.~(\ref{eq:momentum&mass}).
Using the same scaling procedure that led to Eq.~(\ref{appen:eq:function_G}), the particular solutions of interest have the form of
\refstepcounter{equation}
\begin{equation}
u\left ( z,t \right ) =\sqrt{t}f_2(\xi,\eta),\qquad 
\kappa \left ( z,t \right ) = tg_2(\xi,\eta),
\tag{A5a,A5b} \label{appen:eq:solutions_full}
\end{equation}
where $\xi$ remains unchanged and $\eta=\sqrt{t}$.
According to the superposition principle, the solutions for Eqs.~(\ref{appen:eq:no gravity}) are part of the full solutions.
The latter are then written as
\refstepcounter{equation}
\begin{equation}
f_2(\xi,\eta)=f_1(\xi)+\varphi(\eta),\qquad 
g_2(\xi,\eta)=g_1(\xi)+\psi(\eta),
\tag{A6a,A6b} \label{appen:eq:f2&g2}
\end{equation}
where $\varphi$ and $\psi$ represent the additional contributions due to gravity.
Because gravity acts as a uniform body force independent of position, $\varphi$ and $\psi$ have been assumed to depend only on time.
\citet{van2021self} discussed that nonlinear terms in $\varphi$ and $\psi$ would result in singular behavior in the limit $\eta \to 0$, hence linear dependencies are assumed, namely $\varphi=A_1\eta$ and $\psi=C_1\eta$.
Substituting Eqs.~(\ref{appen:eq:solutions_full}) together with Eqs.~(\ref{appen:eq:f2&g2}) into the full governing Eqs.~(\ref{eq:momentum&mass}) yields $A_1=-1/2$ and $C_1=-3B\mathrm{Bo}/4$.
Inserting these coefficients back gives the self-similar solutions presented in Eqs.~(\ref{eq:solutions}).

\nocite{*}

\bibliography{manuscript}% Produces the bibliography via BibTeX.

\end{document}